\documentclass[epj]{webofc}
\usepackage[varg]{txfonts}   
\wocname{EPJ Web of Conferences}
\woctitle{INPC 2013}
\begin{document}
\title{The Proton Form Factor Ratio Measurements at Jefferson Lab}
%
%

\author{Vina Punjabi\inst{1}\fnsep\thanks{\email{vapunjabi@nsu.edu}} \and
        Charles F. Perdrisat\inst{2}\fnsep\thanks{\email{perdrisa@jlab.org}}}

\institute{Norfolk State University, Norfolk, Virginia 23504, USA 
\and
           the College of William and Mary, Williamsburg, Virginia 23187, USA
          }

\abstract
{The ratio of the proton form factors, $G_{Ep}/G_{Mp}$, has been measured from
$Q^2$ of 0.5  GeV$^2$ to 8.5 GeV$^2$, at the Jefferson Laboratory, using the 
polarization transfer
method. This ratio is extracted directly from the measured 
ratio of the transverse and longitudinal polarization components of the recoiling proton 
in elastic electron-proton scattering. The discovery that the proton form factor ratio 
measured in these experiments decreases approximately linearly 
with four-momentum transfer, $Q^2$,  for values above $\approx 1$ GeV$^2$, is one of the most 
significant results to come out of JLab. These results
have had a large impact on progress in hadronic physics; and have 
required a significant rethinking of nucleon structure.
There is an approved experiment at JLab, GEp(5), to continue 
the ratio measurements to 12 
GeV$^2$. A dedicated experimental setup, the Super Bigbite Spectrometer (SBS), will be
built for this purpose. 
In this paper, the present status of the proton elastic 
electromagnetic
form factors and a number of 
theoretical 
approaches to describe nucleon form factors will be discussed. 
  
}
\maketitle
\section{Introduction}
\label{intro}
In three experiments, GEp(1) \cite{jones,punjabi}, GEp(2) \cite{gayou,puck2} and 
GEp(3) \cite{puckett}, in Halls A and C at JLab, the ratio of the proton's
electromagnetic elastic form factors, $G_{Ep}/G_{Mp}$, was measured 
up to four momentum transfer $Q^2$ of 8.5 GeV$^2$ with high precision,
using the recoil polarization technique. The initial discovery that the proton
form factor ratio measured in these three experiments decreases 
approximately linearly with four-momentum transfer, $Q^2$, for values 
above $\approx 1$ GeV$^2$, was modified by the GEp(3) results,
which suggests a slowing down of this decrease. 

Use of the double-polarization technique to 
obtain the elastic nucleon form factors has resulted 
in a dramatic improvement of the quality of two of the four nucleon electromagnetic
form factors, $G_{Ep}$ and $G_{En}$. It has also changed our understanding of the proton 
structure, having resulted in a distinctly different $Q^2$- dependence for
both $G_{Ep}$ and $G_{Mp}$,
contradicting the prevailing wisdom of the 1990's based on
cross section measurements, namely that 
$G_{Ep}$ and $G_{Mp}$ obey a ``scaling'' relation
$\mu G_{Ep}\sim G_{Mp}$. A related consequence  
of the faster decrease of $G_{Ep}$ revealed by
the Jefferson Lab (JLab) polarization results was the disappearance of the early 
scaling of $F_2/F_1 \sim 1/Q^2$ predicted by perturbative QCD. 

\section{Elastic {\bf ep} cross section and form factors}

In terms of the electric and magnetic Sachs form factors, $G_E$ and $G_M$, the lab cross section for 
elastic {\it ep} scattering can be written as:
\begin{equation}
\frac{d\sigma}{d\Omega}=\frac{d\sigma}{d\Omega}_{Mott}\frac{G_{Ep}^2+\frac{\tau}{\varepsilon}G_{Mp}^2}{1+\tau},
\label{eq:dsigma}
\end{equation}
\noindent
where where $\tau = \frac{Q^2}{4m_p^2}$ is the dimensionless 4-momentum transfer squared, and
$\varepsilon$ is the longitudinal polarization of the virtual
photon, $\varepsilon =[1+2(1+\tau)\tan^2\frac{\theta_e}{2}]^{-1}$; $\theta_e$ is the lab
electron scattering angle.
Equation \ref{eq:dsigma} leads to a simple separation method for $G_{E_{p}}^2$ and $G_{M_{p}}^2$, referred to as
Rosenbluth (or LT) separation \cite{rosen}. 

\begin{figure}[ht]
\begin{minipage}[b]{0.45\linewidth}
\begin{center}
\vspace{0.3in}
\includegraphics[width=55mm,angle=90]{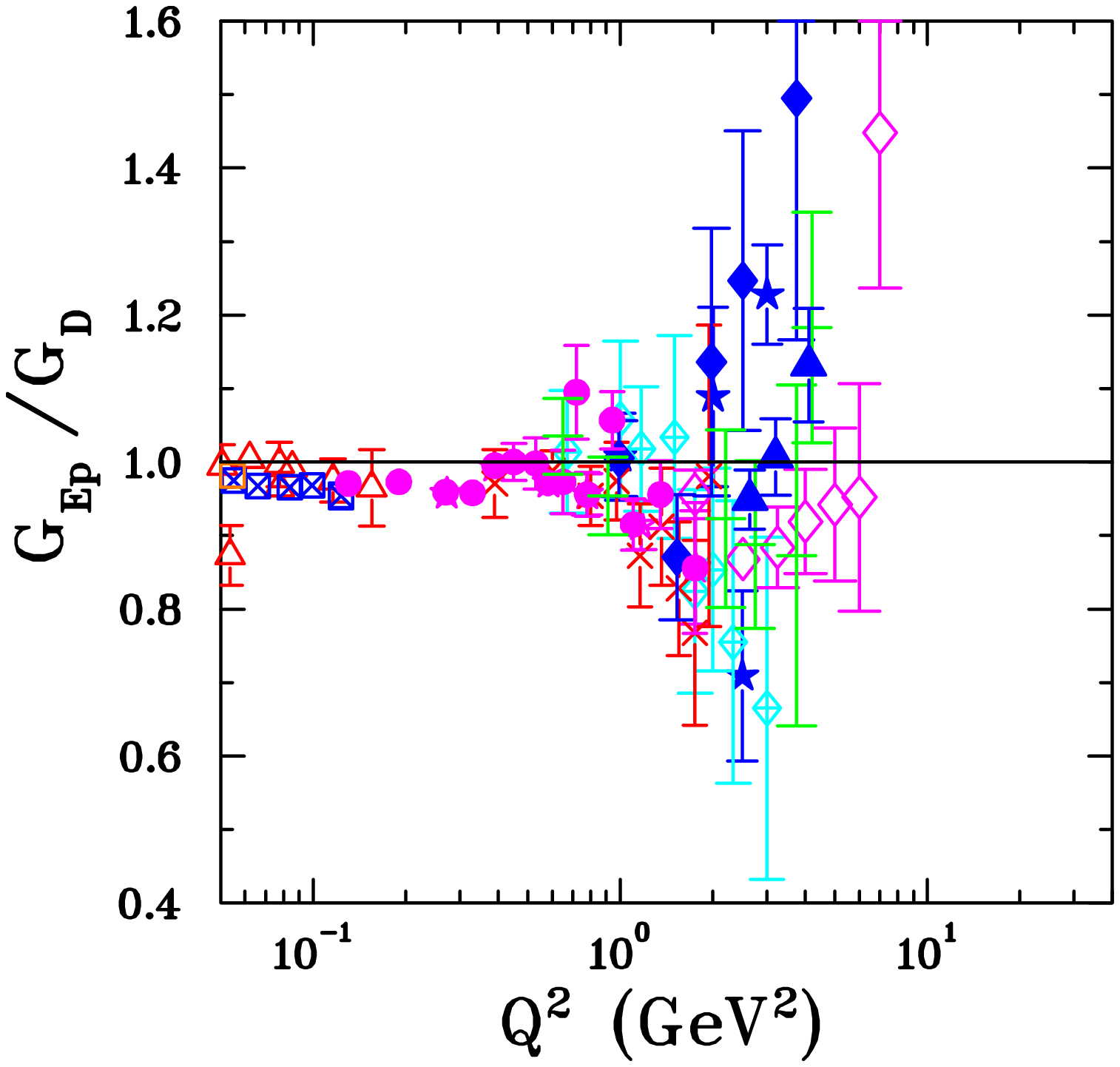}
\caption{World data base for $G_{E{p}}$ obtained by the Rosenbluth method.}
\label{fig:gepgd}
\end{center}
\end{minipage}\hfill
\begin{minipage}[b]{0.45\linewidth}
\begin{center}
\includegraphics[width=55mm,angle=90]{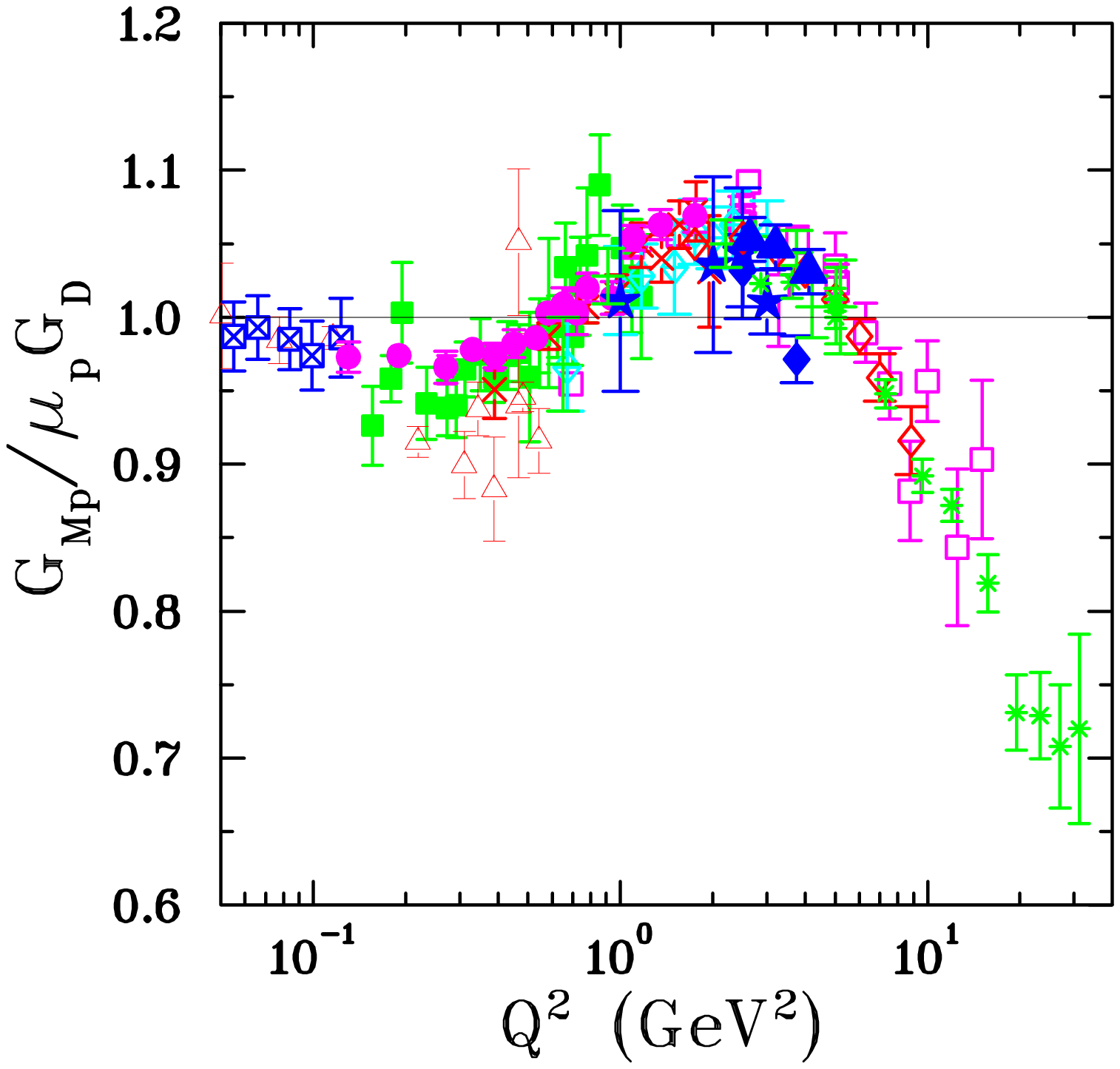}
\caption{World data base for $G_{M{p}}$ obtained by the Rosenbluth method.}
\label{fig:gmpgd}
\end{center}
\end{minipage}\hfill
\end{figure}

In the Rosenbluth method, the separation of $G_{Ep}^2$ 
and $G_{Mp}^2$ is achieved by fitting data with a straight line fit at a given Q$^2$ over a range of 
$\epsilon$ obtained by changing the beam energy, $E_e$ and electron scattering angle, $\theta_e$.
The form factors, $G_{Ep}$ and $G_{Mp}$, obtained from all cross section measurements 
are shown in Fig. \ref{fig:gepgd} and \ref{fig:gmpgd}; they have been divided by the dipole form factor 
$G_D=(1+\frac{Q^2}{0.71})^{-2}$. Evidently the form factors divided by $G_D$ appear to 
remain close to $1$. This behavior suggested that 
$G_{Ep}$, and $G_{Mp}$ have similar spatial distributions.
\vspace{-0.2in} 

\vspace{-0.35in}

\section{Recoil Polarization Method}

The relationship between the Sachs electromagnetic form factors and
the polarization transfer to the recoil proton in $^1H(\vec{e}, e^{\prime}
\vec{p} \,)$ scattering was first developed by Akhiezer and Rekalo
\cite{akhiezer}, and later discussed in more detail by Arnold, Carlson,
and Gross \cite{arnold}.
For single photon exchange, the 3 components of the transferred polarization are:

\vspace{-0.1in}

\begin{center}
\begin{eqnarray}
\label{eq:transpol}
P_n & = & 0 \\
\label{eq:transpol2}
 h P_eP_{\ell} & = &  h P_e \left( \frac{E_e + E_e^\prime}{m_p} \right) \frac{\sqrt{\tau(1 + \tau)} \, 
G_{M p}^2(Q^2) 
\, \tan^2\frac{\theta_e}{2}}{G_{E p}^2(Q^2) + \frac{\tau}{\epsilon} G_{M p}^2(Q^2)}  \\
\label{eq:transpol3}
 h P_e P_t & = &  h P_e \frac{2 \sqrt{\tau(1 + \tau)} \, {G_{E p} \, G_{M p} \, \tan\frac{\theta_e}
{2}}}{G_{E p}^2(Q^2) + \frac{\tau}{\epsilon} G_{M p}^2(Q^2)}
\end{eqnarray}
\end{center}
\noindent for the normal, in-plane longitudinal and transverse polarization components $P_n$, $P_{\ell}$
 and $P_t$, respectively; the $h=\pm$ stands for the electron beam helicity, and $P_e$ for the electron beam polarization.  

For each $Q^2$, a single measurement of the azimuthal angular
distribution of the proton scattered in a secondary target 
gives both the longitudinal and transverse polarizations.
Combining Eqs. \ref{eq:transpol2} and \ref{eq:transpol3} directly provides:
\begin{equation}
\label{eq:ugegm}
\frac{G_{E p}}{G_{M p}} = -\frac{P_t}{P_{\ell}} \frac{(E_e + E_e^\prime)}{2 m_p} \tan 
\frac{\theta_e}{2};
\end{equation}

\begin{figure}[hbt]
\begin{minipage}[b]{0.45\linewidth}
\begin{center}
\includegraphics[width=50mm,angle=90]{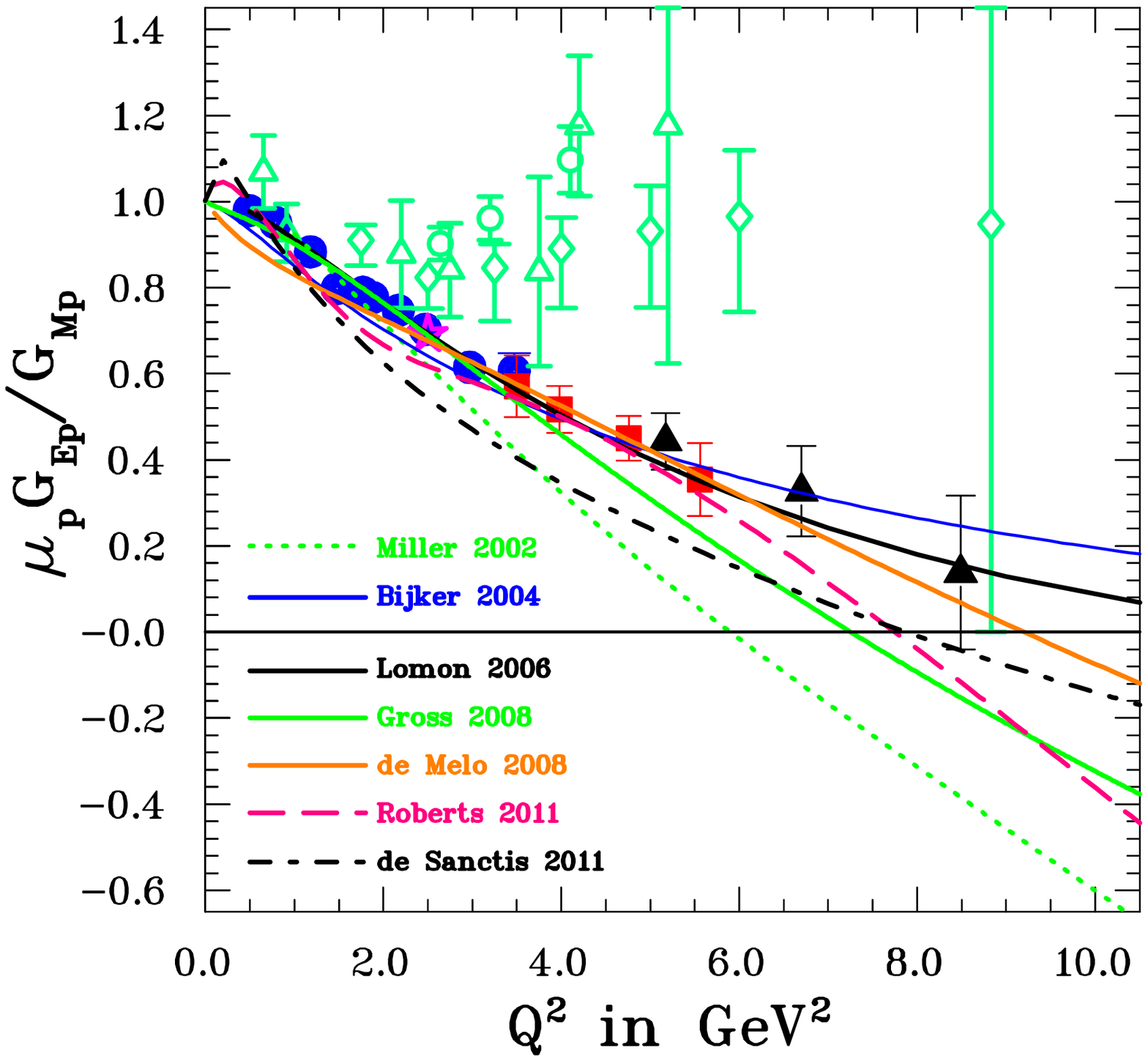}
\caption{\small{The ratio $\mu_pG_{E_p}/G_{M_p}$ obtained in polarization
 transfer experiments shown as filled symbols, refs. \cite{jones,gayou,punjabi,
puckett,meziane,puck2}. Empty symbols: Rosenbluth results
of refs. ~\cite{andivahis,christy,qattan}. }} 
\label{fig:gepgmp}
\end{center}
\end{minipage}\hfill
\begin{minipage}[b]{0.45\linewidth}
\begin{center}
\includegraphics[width=50mm,angle=90]{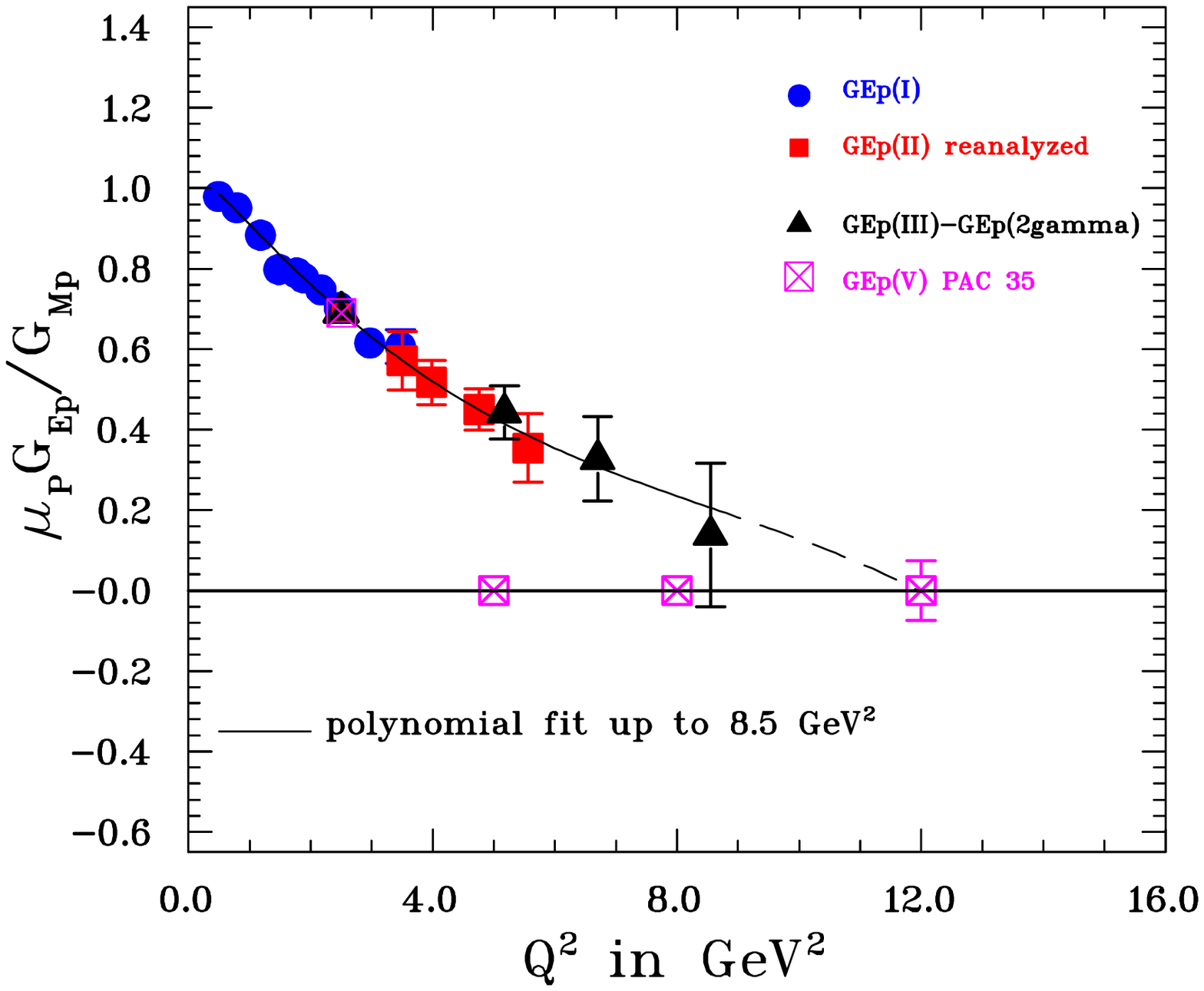}
\caption{\small{The three data points to be obtained in the GEp(5) experiment \cite{GEp5}. The solid line is a polynomial fit to the existing data, extrapolated as dashed line.}}
\label{fig:GEpV}
\end{center}
\end{minipage}
\end{figure}

The striking disagreement of the polarization data with the Rosenbluth results is
illustrated in Fig. \ref{fig:gepgmp}; the Refs. to the various curves are 
\cite{gamiller,bijker,lomon,gross,demelo,cloet,desanctis}.

\vspace{-0.25in}

\section{The Super Bigbite Spectrometer}

Measurements of $G_{E p}/G_{M p}$ and $G_{E n}$ from elastic {\bf ep} or {\bf en} scattering, to yet 
larger Q$^2$, requires a new approach, largely because of the rapid decrease of both the elastic
scattering cross sections and of the pCH$_2$ analyzing power $A_y$ with increasing Q$^2$. 

The approved Super Bigbite Spectrometer project at JLab Hall A \cite{sbs}, or SBS was
originally conceived to allow measurement
of $G_{E p}/G_{M p}$ to up to 15 GeV$^2$ (GEp(5))\cite{GEp5}. 
On the proton side it consists of a dipole with 
horizontal B-field and very large aperture. The resulting high flux of low energy photons will be
overcome by using GEM detectors. For GEp(5) (currently approved to 12 GeV$^2$), a double polarimeter 
with 2 CH$_2$ analyzers and 2 GEM detector clusters, 
is being built. The trigger for the proton arm will be generated in the hadron calorimeter downstream
from the polarimeter. 

The electron  will be detected in a new electromagnetic calorimeter, preceded by a GEM coordinate detector.   
The anticipated error bars and Q$^2$ values are shown in Fig. \ref{fig:GEpV}.     
 
\section {Conclusions}

Much has happened since the results of the first JLab {\bf ep} form factor double-spin experiment, which
challenged 50 years of cross section measurements. A recent measurement of $G_{E p}/G_{M p}$
has reached the maximum Q$^2$ value of 8.5 GeV$^2$, and indicates that the quasi-linear decrease
of this ratio most likely comes to an end in this range of Q$^2$-values. 

The high-Q$^2$ surprise in $G_{Ep}/G_{Mp}$, have led to a several fundamental changes in the picture
of the internal structure of the proton, and a revival of interest for elastic form factor  data
among nuclear theorists.

The recent results from double polarization experiments for the proton, together with the anticipated results
following the 12 GeV upgrade of the JLab accelerator, will provide answers to a number of open questions
crucial to the understanding of fundamental properties of the proton, and the nature of QCD in the confinement regime.


\section{Acknowledgments}
The authors acknowledge grant support from the U.S. Department of Energy, DE-FG02-89ER40525 (VP) and 
the National Science Foundation, PHY1066374 (CFP).


\begin{thebibliography}{}

\bibitem{jones} M.K. Jones {\it et al.}, Phys. Rev. Lett. {\bf 84}, 1398 (2000).

\bibitem{punjabi} V. Punjabi {\it et al.} Phys. Rev. C {\bf 71} (2005) 055202.

\bibitem{gayou} O. Gayou {\it et al.}, Phys. Rev. Lett. {\bf 88}, 092301 (2002).

\bibitem{puck2} A.J.R Puckett {\it et al}, Phys. Rev. {\bf 85}, 045203 (2012).

\bibitem{puckett} A.J.R Puckett {\it et al}, Phys. Rev. Lett.  {\bf 104}, 242301 (2010).

\bibitem{rosen} M.N. Rosenbluth, Phys. Rev. {\bf 79}, 615 (1950).

\bibitem{akhiezer} A.~I.~Akhiezer and M.~P.~Rekalo, Sov.\ J.\ Part.\ Nucl.\  {\bf 4}, 277 (1974).

\bibitem{arnold} R.G. Arnold, C.E. Carlson, F. Gross, Phys. Rev. C  {\bf 23}, 363 (1981).

\bibitem{gamiller} M.R. Frank, B.K. Jennings, and G.A. Miller, Phys. Rev. C {\bf 54}, 920 (1996); 
                   G.A. Miller and M.R. Frank, Phys. Rev. C {\bf 65}, 065205 (2002).
\bibitem{bijker} R. Bijker and F. Iachello,  Phys.Rev. C {\bf 69} (2004) 068201.

\bibitem{lomon} E.L. Lomon, nucl-th/0609029v2.

\bibitem{gross} F. Gross and P.~Agbakpe, Phys. Rev.  C {\bf 73}, 015203 (2006).
  
\bibitem{demelo} J.P.B.~de~Melo, T.~Frederico, E.~Pace, S.~Pisano and G.~Salme, Phys. Lett. B {\bf 671}, 153 (2009).

\bibitem{cloet} I.C.~Clo\"{e}t and C.D.~Roberts, Proc. of Sc. LC2008:047 (2008); 
arXiv:0811.2018 [nucl-th] (2008); I.C.~Clo\"{e}t, G.~Eichmann, B.~El-Bennich, T.~Kl\"{a}hn and 
C.D.~Roberts, Few Body Syst. {\bf 46}, 1 (2009)

\bibitem{desanctis} M. de Sanctis {\it et al}, Phys. Rev. C {\bf 76}, 062201 (2007).

\bibitem{meziane} M.~Meziane {\it et al.} [GEp2gamma Collaboration], Phys. Rev. Lett. {\bf 106}, 132501 (2011)

\bibitem{andivahis} L. Andivahis {\it et al.}, Phys. Rev. D {\bf 50} 5491 (1994).

\bibitem{christy} M.E.Christy  {\it et al.}, Phys. Rev. {\bf C70} 015206 (2004).

\bibitem{qattan} I.A. Qatan  {\it et al.}, Phys. Rev. Lett. {\bf 94} 142301 (2005).

\bibitem{GEp5}http://hallaweb.jlab.org/collab/PAC/PAC32/PR12-07-109-Ratio.pdf

\bibitem{sbs} http://hallaweb.jlab.org/12GeV/SuperBigBite/ 

\end{thebibliography}
\end{document}